\documentclass[%
 reprint,
superscriptaddress,
 amsmath,amssymb,
 aps,
]{revtex4-2}

\usepackage{amsmath}
\usepackage{amssymb}
\usepackage{amsfonts}
\usepackage{braket}
\usepackage{wrapfig}
\usepackage{float}
\usepackage{caption}
\usepackage{subcaption}
\usepackage{amsthm}
\usepackage{float}
\usepackage{stfloats}
\usepackage{graphicx}
\usepackage{dcolumn}
\usepackage{bm}

\begin{document}

\preprint{APS/123-QED}

\title{Quantum and Non-local Effects Offer LiDAR over 40dB Advantage}

\author{$\text{Phillip S. Blakey}$}
\affiliation{Edward S. Rogers Faculty of Applied Science and Engineering, University of Toronto}
\email[email: ]{phil.blakey@mail.utoronto.ca}
\author{\text{Han Liu}}
\affiliation{Edward S. Rogers Faculty of Applied Science and Engineering, University of Toronto}
\author{\text{Georgios Papangelakis}}
\affiliation{Edward S. Rogers Faculty of Applied Science and Engineering, University of Toronto}
\author{Yutian Zhang}
\affiliation{Edward S. Rogers Faculty of Applied Science and Engineering, University of Toronto}
\author{Zacharie M. Léger}
\affiliation{Edward S. Rogers Faculty of Applied Science and Engineering, University of Toronto}
\author{Meng Lon Iu}
\affiliation{Edward S. Rogers Faculty of Applied Science and Engineering, University of Toronto}
\author{Amr S. Helmy}
\affiliation{Edward S. Rogers Faculty of Applied Science and Engineering, University of Toronto}
\begin{abstract}
Non-local effects have the potential to radically move forward quantum enhanced LiDAR to provide an advantage over classical LiDAR not only in laboratory environments but practical implementation. In this work, we demonstrate a 43dB lower signal-to-noise ratio using a quantum enhanced LiDAR based on time-frequency entanglement compared with a classical phase-insensitive LiDAR system. Our system can tolerate more than 3 orders of magnitude higher noise than classical single-photon counting LiDAR systems before detector saturation. To achieve these advantages, we use non-local cancellation of dispersion to take advantage of the strong temporal correlations in photon pairs in spite of the orders of magnitude larger detector temporal uncertainty. We go on to incorporate this scheme with purpose-built scanning collection optics to image non-reflecting targets in an environment with noise.

\end{abstract}


\maketitle


\section{\label{sec:level1}Introduction:\protect\\ }
Quantum photonic technologies have been gaining significant momentum, where generation \cite{kang2016monolithic}, detection, processing, and utilization of quantum states have been advanced significantly towards practical implementation and integration \cite{moody2021roadmap}. This, in turn, offers numerous possibilities for their usage in advancing well-established fields, including sensing, communications, and computing. 

In parallel with such advances, several emerging applications including LiDAR \cite{rogers2021universal} and biomedical imaging \cite{taylor2016quantum} have been encountering challenges to meet their noise resilience roadmap goals using practical, robust, and low complexity classical protocols. The recent advances in quantum photonic technologies offer a promising route to addressing many of these challenges. 

Quantum Illumination (QI) has been proposed, investigated, and demonstrated over the past decade as a solution to the challenges of combating environmental noise in LiDAR and other imaging applications \cite{lloyd2008enhanced,tan2008quantum}. When compared with optimal classical detection using a coherent state, QI has been shown to provide the largest improvement theoretically possible\cite{guha2009gaussian}. However, to date, implementations of QI have been unable to reach such theoretically predicted bounds irrespective of the approach taken \cite{zhang2015entanglement}. For both QI and optimal classical coherent detection, the states used must have a stable phase. Many practical limitations make it extremely difficult to keep the interacting waves phase-locked, even with the most complex stabilization systems. As a consequence, optimal coherent states are not usually used in practical classical LiDAR applications, where intensity based detection schemes are more convenient \cite{mcmanamon2012review}, and therefore using such states as a benchmark may not be representative.

Alternative phase-insensitive quantum enhanced target detection schemes have been reported using quantum correlations, including those in both intensity and time \cite{england2019quantum,liu2019enhancing}. Such phase-insensitive systems provide dramatic simplification in implementation, and are attractive for systems where large phase noise is introduced during operation. The quantum enhancement has been shown when compared to phase-insensitive classical states \cite{liu2020target, he2020non};
however, an improved performance has been demonstrated over classical phase-insensitive counterparts only in a limited range of noise powers. This has been due to the saturation of detectors at high noise levels, where the most significant advantage of such schemes is obtained.  Further, the use of strong temporal correlations is limited by the relatively large detector time uncertainty. This uncertainty effectively erases any correlation shorter than the detector time uncertainty, reducing the advantages offered by temporal correlation (Supplimentary Material). For example, commercial superconducting nanowire single photon detectors have time uncertainty $\simeq 50$ps which is several orders of magnitude larger than the shortest achievable correlation times \cite{abolghasem2009bandwidth}. State of the art detectors have time uncertainty $\simeq 3$ps \cite{korzh2020demonstration}, however this comes at the cost of reduced efficiency while still being several orders of magnitude larger than the shortest correlation times.  

In this work we utilize quantum temporal correlation to assist in the discrimination of a target from the background noise for LiDAR application. By measuring in a rotated basis between time and frequency---the Fractional Fourier domain---we can magnify the probe-reference temporal uncertainty while maintaining the same degree of correlation. This allows us to fully use the probe-reference correlations to distinguish a target from background noise. The uncorrelated noise is broadened well beyond the detector uncertainty. Applying a suitable temporal window can then filter the noise that no longer overlaps the signal. With this method, we were able to enhance the signal to noise ratio by up to $43.1$dB compared with a phase-insensitive classical target detection counterpart using the same probe power. This method retains the ease of implementation of previously mentioned target detection schemes, while also increasing the noise power that can be tolerated before detector saturation.

We then integrate this scheme with a purpose built telescope to image a target in a highly noisy environment thus demonstrating the applicability of this technique. The telescope allows for a wide scanning angle while maintaining a superior coupling efficiency into a single-mode fiber, when compared to alternative telescope designs.

\section{System Description}

\begin{figure}[H]
    \centering
    \includegraphics[width=1.0\textwidth]{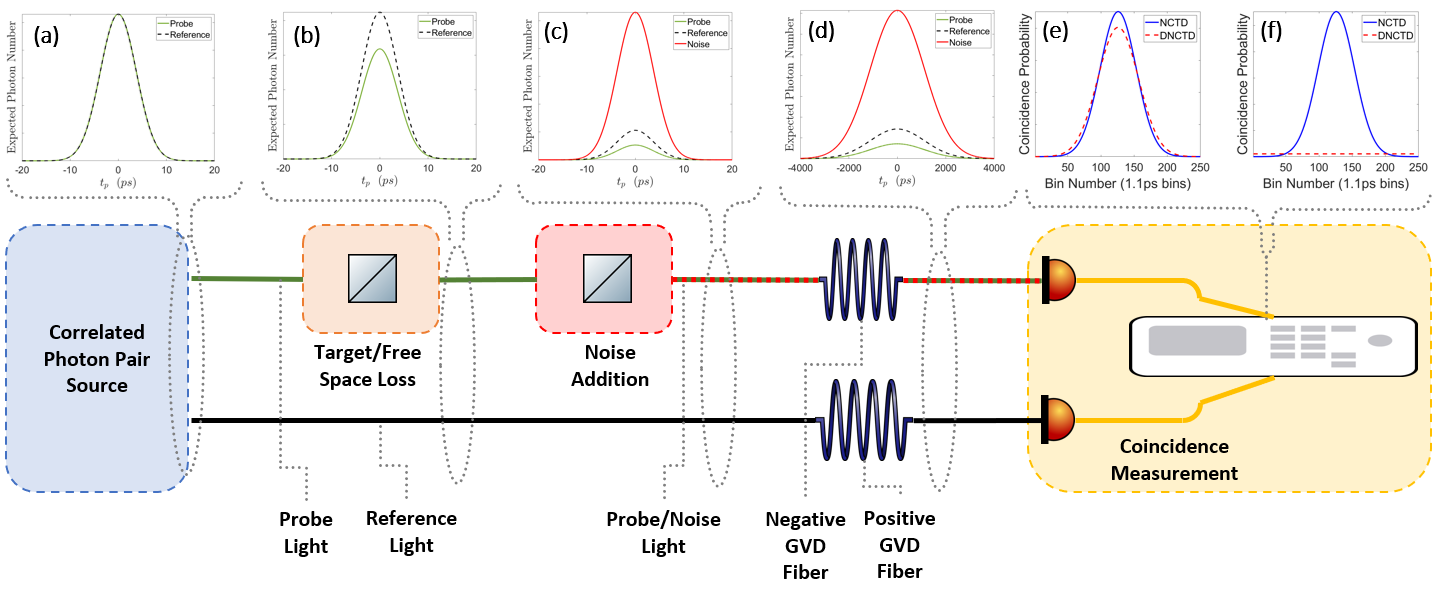}
    \caption{(a) Temporal distribution of probe and reference photons directly after SPDC. (b) Temporal distribution of the probe and reference photons after free-space/target loss. (c) Temporal distribution of the noise, probe, and reference photons after coupling of environmental noise. (d) Temporal distribution of the noise, probe, and reference photons after dispersion. (f) Theoretical coincidence probability histograms for true coincidences (e) and false coincidence (f) in both the non-dispersed (blue) and dispersed (red) regimes. The model uses JSA bandwidths of 100fs and 17.7ps FWHM estimated from the Ti:Sapphire laser and SHG spectrum of the PPLN waveguide, the dispersion and length of the fibers (18$\text{ps}/ \text{nm}\cdot \text{km}$, 5km), and detector time uncertainty (83.3ps) FWHM.}
    \label{fig:Logical Diagram}
\end{figure}

The underlying working principles of our system are illustrated in figure \ref{fig:Logical Diagram}. First, non-classical temporally correlated photon pairs are generated through femtosecond pumped spontaneous parametric down-conversion (SPDC). The pump is chosen to be sufficiently weak such that the expected number of probe and reference photons is much less than one.

The probe photon is sent out into the environment while the reference photon is stored locally. The probe photon incurs loss during propagation towards and returning from the target, reducing the expected photon number in the probe beam (figure \ref{fig:Logical Diagram}b). During the probe photon's propagation, environmental noise is coupled into the probe path. Working under the assumption that the noise has the same spectral/temporal distribution as the probe photon (figure \ref{fig:Logical Diagram}c) (as other noise may be filtered out classically), we apply anomalous dispersion to the probe/noise photon. This broadens the temporal distribution of the probe/noise photon, resulting in a lower probability of finding a photon in a finite time window. Normal dispersion is then applied to the reference photon which also broadens the temporal distribution of the reference photon (figure \ref{fig:Logical Diagram}d). A coincidence measurement is then performed on the two paths. Due to the quantum correlations between the probe and reference photons, the effects of the dispersion are cancelled and the true coincidence peak appears almost as though the photons were not dispersed \cite{franson1992nonlocal,baek2009nonlocal}. Conversely, the noise and reference photons only share classical correlations and so the effects of dispersion cause a broadening of the coincidence peak (figure \ref{fig:Logical Diagram}f). By choosing an appropriate temporal window, the probability of measuring a false coincidence between a noise photon and a reference photon is reduced, while the probability of measuring a true coincidence between a probe photon and a reference photon is essentially unchanged.

\onecolumngrid

\twocolumngrid

\begin{figure}[H]
    \begin{subfigure}{0.49\columnwidth}
  \centering
  \includegraphics[width=\columnwidth]{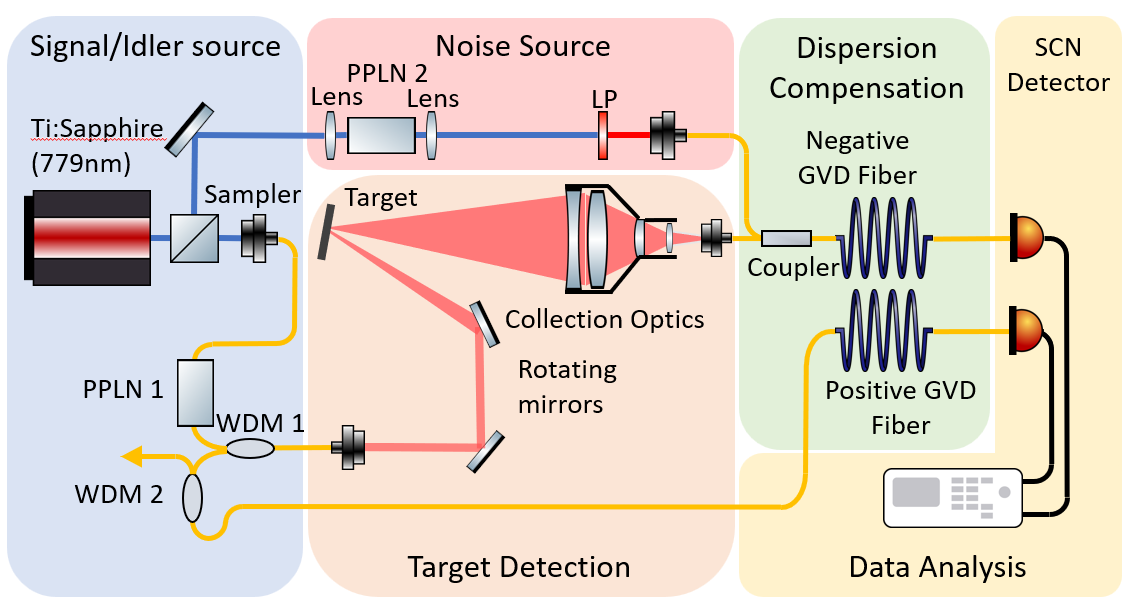}
  \caption{}
  \label{fig:Schematic}
\end{subfigure}
\begin{subfigure}{0.49\columnwidth}
      \centering
    \includegraphics[width=\columnwidth]{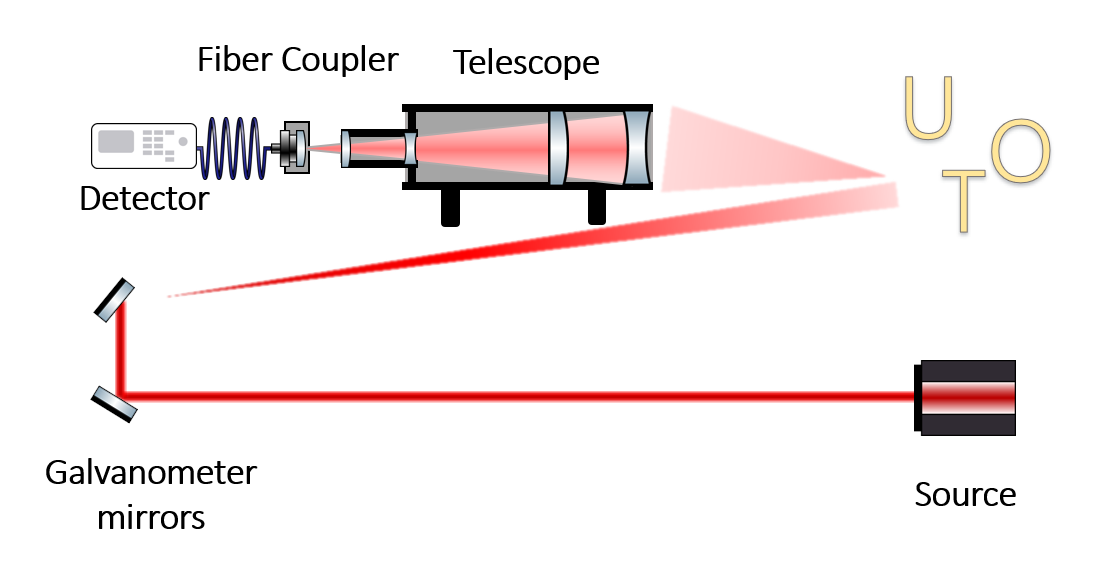}
    \caption{}
    \label{fig:Scanning Setup}
\end{subfigure}
\caption{(a) Experimental schematic for the DNCTD protocol including scanning and collection optics. (b) Purpose built scanning and collection optics for LiDAR demonstration.}
\end{figure}

To demonstrate LiDAR capabilities, an apparatus for 3D imaging was designed specifically for use with single mode fiber coupled superconducting nanowire detectors (figure \ref{fig:Scanning Setup}). First, the probe photons from the SPDC source are collimated onto a pair of galvanometer mirrors. These mirrors direct the probe photon onto a target in the field of view of a telescope. The rotation of the mirrors allows for scanning of the target in both x and y directions. To reduce the mode mismatch between the collected light and SMF, a negative meniscus lens was used at the telescope aperture to reduce the angle offset. Using the constant velocity of the probe photons in addition with the time delay between probe and reference photons, the depth of a target can be resolved allowing for 3D imaging.
\newline\newline

\onecolumngrid

\twocolumngrid

\section{System Design}

Photon pairs generated through SPDC exhibit entanglement in the time-frequency degree of freedom. This can manifest itself as a correlation in detection time of the two photons at separate detectors. This correlation can aid in distinguishing probe light from background noise that has been coupled into the collection optics. Due to the relatively large detector time uncertainty, the full advantage of the temporal correlations cannot be used. It is beneficial then, to measure in a different basis between time and frequency, trading some of the temporal correlation for frequency correlation. In our setup, this is achieved by applying negative dispersion to the probe/noise photon and positive dispersion to the reference photon. The measurable effect of this can be seen in figure (\ref{fig:Logical Diagram} e) and (\ref{fig:Logical Diagram} f) as a slight broadening of the true coincidence peak due to the non-maximal probe-reference entanglement and a substantial broadening of the noise peak. This allows for filtering of the previously indistinguishable un-correlated noise increasing the SNR. The mathematical details of this improvement are described in the supplementary material. To compare the performance of the new scheme, which we call dispersed non-classical target detection (DNCTD), with previous ones, we also measure the SNR in a classical target detection (CTD) analog, as well as a coincidence based non-classical target detection (NCTD). The SNRs for these other two schemes are given by
\begin{align}
    SNR_{CTD} &= \frac{\nu \tau_p}{N}\label{CTD SNR},\\
    SNR_{NCTD} &= \frac{\nu\tau_p\tau_r}{N\nu\tau_r}\label{NCTD SNR},
\end{align}
where $\nu$ is the pair generation rate, $\tau_p$ and $ \tau_r$ are the probe and reference path transmissions, and $N$ is the noise singles rate. The improvement from CTD to NCTD is then given by 
\begin{align}\label{NCTD Improvement}
    SNR_{NCTD} - SNR_{CTD} &= \frac{1}{\nu}. 
\end{align}
To compare the CTD, NCTD, and DNCTD schemes we must define the SNR as a function of some variable that is independent of additional loss after the noise and probe are combined. To achieve this we define the normalized noise power and normalize probe power to be $\frac{N}{\nu\tau_p}$ and $\frac{\nu \tau_p}{N}$ respectively.

Our scheme differs from QI as demonstrated in \cite{zhang2015entanglement} in both implementation and assumptions about the environment. To develop a easily implementable LIDAR setup we restrict ourselves to phase-insensitive detection methods for both our scheme and the classical scheme we compare with. Further, we do not take the background noise to be thermal radiation but assume that we can optimally filter the coupled noise, removing anything that is not spectrally identical to the probe photon.

\section{Experimental Results}
\begin{figure}
\begin{subfigure}{0.49\columnwidth}
  \centering
  \includegraphics[width=\columnwidth]{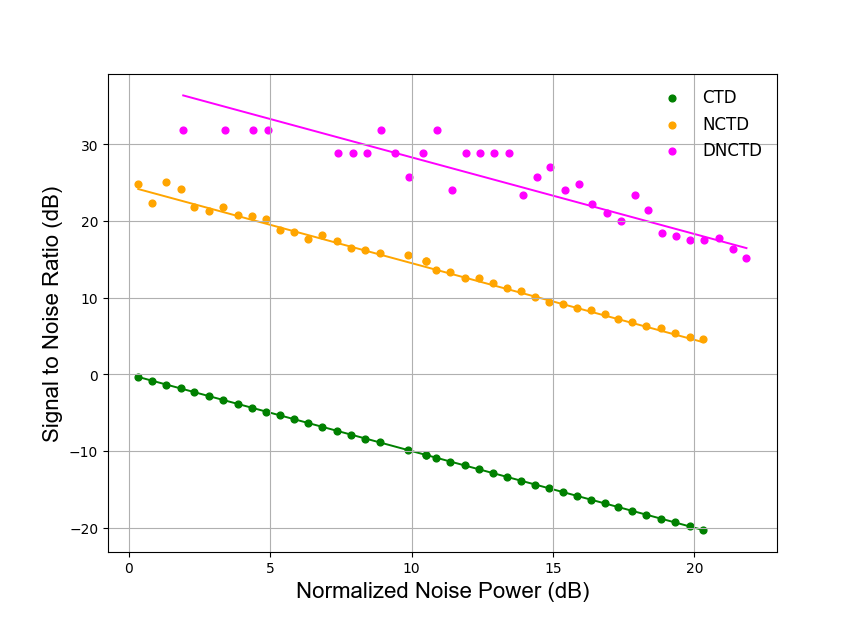}
  \caption{}
  \label{fig:varying noise}
\end{subfigure}
\begin{subfigure}{0.49\columnwidth}
  \centering
  \includegraphics[width=\columnwidth]{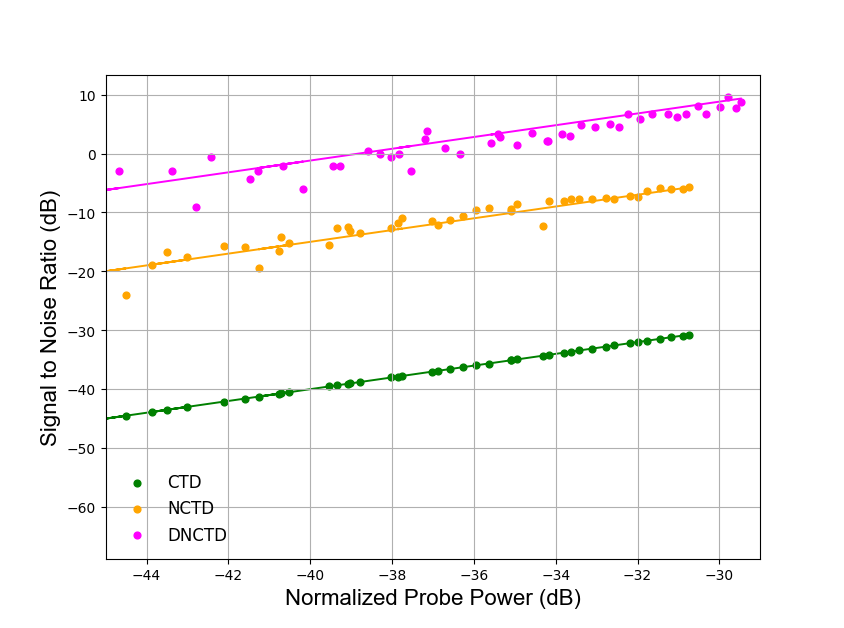}
  \caption{}
  \label{fig:varying probe}
\end{subfigure}
\begin{subfigure}{0.49\columnwidth}
  \centering
  \includegraphics[width=\columnwidth]{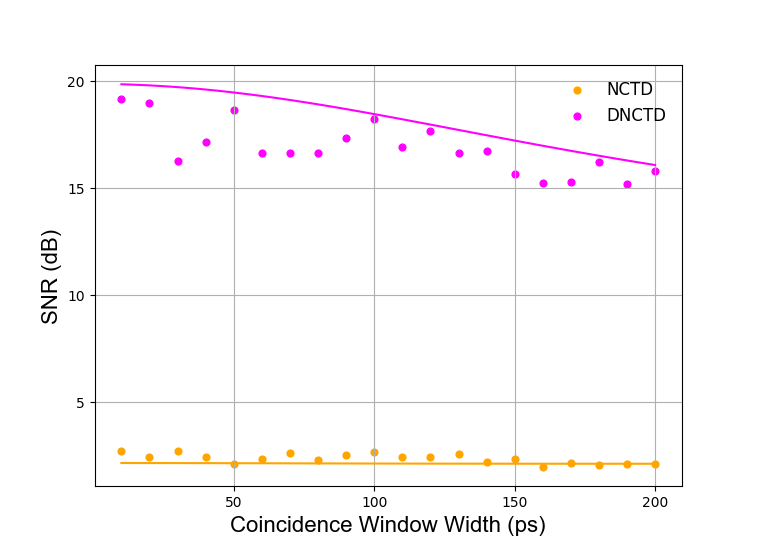}  
  \caption{}
  \label{fig:Coinc Window Data}
\end{subfigure}
\begin{subfigure}{0.49\columnwidth}
  \centering
  \includegraphics[width=\columnwidth]{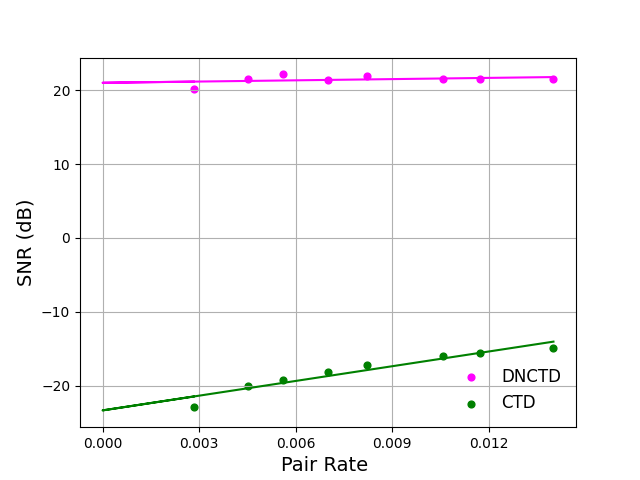}  
  \caption{}
\label{fig:Varying Power}
\end{subfigure}
\label{fig:Data Figure}
\caption{(a) Signal to Noise ratio for varying noise power. (b) Signal to Noise ratio for varying probe power. (c) SNR as a function of coincidence window width for DNCTD (magenta) and NCTD (orange). (d) SNR for DNCTD and CTD as a function of SPDC pair rate. The coloured lines depict the theoretical predictions for comparison with experimental data.}
\end{figure}

The experimental setup used to demonstrate the performance of the DNCTD scheme compared with the NCTD and CTD schemes is shown in figure \ref{fig:Schematic} and examined in detail in the supplementary material. For this comparison, we examine the SNR of each scheme for increasing environmental noise (\ref{fig:varying noise}), increasing probe channel loss (\ref{fig:varying probe}), coincidence window width \ref{fig:Coinc Window Data}, and pump power (\ref{fig:Varying Power}) bypassing the target scanning and collection optics.

First, the normalized noise power was varied while keeping the remaining experimental parameters (pair rate, coincidence window, probe/reference transmission) fixed (figure \ref{fig:varying noise}) in a regime convenient for this measurement. The DNCTD SNR is a roughly constant $14.1$dB higher than the NCTD SNR, and the NCTD is a roughly constant $24.5$dB higher than CTD SNR. All three schemes exhibit a approximately linear relationship with the normalized noise power in good agreement with theoretical predictions (coloured lines) calculated through equations \eqref{CTD SNR},\eqref{NCTD SNR}, and the DNCTD theory (Supplementary material). 

The SNR was then measured as a function of normalized probe power for the CTD, NCTD and DNCTD (figure \ref{fig:varying probe}) with unique fixed experimental parameters (pair rate, probe/reference transmission, noise power, and coincidence window). The improvement from the CTD to NCTD schemes was calculated from the measured data through equation \eqref{NCTD Improvement} to be an approximately constant $26.0$dB and the improvement from NCTD to DNCTD was measured to be around a constant $12.7dB$ yielding a total improvement of $38.7$dB. The SNR exhibits a nearly linear relationship with the normalized probe power for all three schemes as predicted by equations \eqref{CTD SNR} and \eqref{NCTD SNR}, and the DNCTD theory with the improvements form CTD to NCTD and NCTD to DNCTD remaining approximately constant for all probe powers as predicted.

\newpage
\onecolumngrid

\twocolumngrid

The SNR was also measured for varying coincidence window widths from 10ps to 200ps (figure \ref{fig:Coinc Window Data}). For this measurement the noise power was set as high as possible given the VOA configuration (Supplementary material) and the pair rate was set large enough to obtain a stable curve at small coincidence windows over a 100s integration time. Using these parameters, DNCTD can achieve an additional ~3.78dB improvement over the NCTD scheme by changing the coincidence window from 200ps to 10ps. While reducing the coincidence window below the detector uncertainty reduces the number of measured coincidences, this has a more significant impact on the dispersed noise coincidences than the true coincidences resulting in a higher SNR. This can be understood from figure (\ref{fig:Logical Diagram}f) where the noise distribution is essentially flat in time (for DNCTD) while remaining in a comparatively sharp Gaussian peak for the NCTD scheme. Theoretical calculations performed by integrating the coincidence probabilities (\ref{fig:Logical Diagram}f) over different coincidence window widths are plotted along side the measured data and are in good agreement with the measured results.

Finally, we attempt to show the maximum SNR improvement from CTD to DNCTD possible for our system by varying the pump power, quantified in terms of the SPDC pair generation rate $\nu$ (\ref{fig:Varying Power}). The noise power was set as high as possible to achieve the highest SNR difference. The highest measured SNR improvement is $43.1$dB, however much higher improvements are measurable given a higher noise power or longer detection time to adequately estimate the number of noise coincidence counts which become increasingly small at low pair rates. As predicted by equation \eqref{NCTD SNR} in conjunction with the advantage dispersion compensation provides being independent of pair rate, the measured DNCTD SNR is essentially independent of pair rate remaining at $21.45\pm 1.25$dB for all measured pair rates. Since the improvement from CTD to NCTD is given by the inverse of the pair rate, the overall improvement from CTD to DNCTD is made largest for minimal pump power, limited only by the system loss and detection efficiency causing the coincidence counts to vanish. Since coincidence counts remain above the noise floor at pair rates where the singles rates are well below the detector dark counts, we must infer the CTD SNR by a linear fit as a function of pair rate.

\begin{figure}
\begin{subfigure}{0.49\columnwidth}
  \centering
  \includegraphics[width=.9\columnwidth]{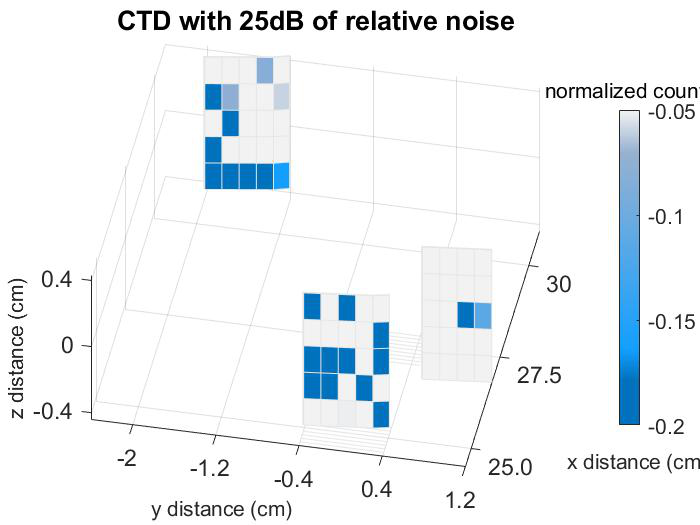}
  \caption{}
  \label{fig:CTD Noise}
\end{subfigure}
\begin{subfigure}{0.49\columnwidth}
  \centering
  \includegraphics[width=.9\columnwidth]{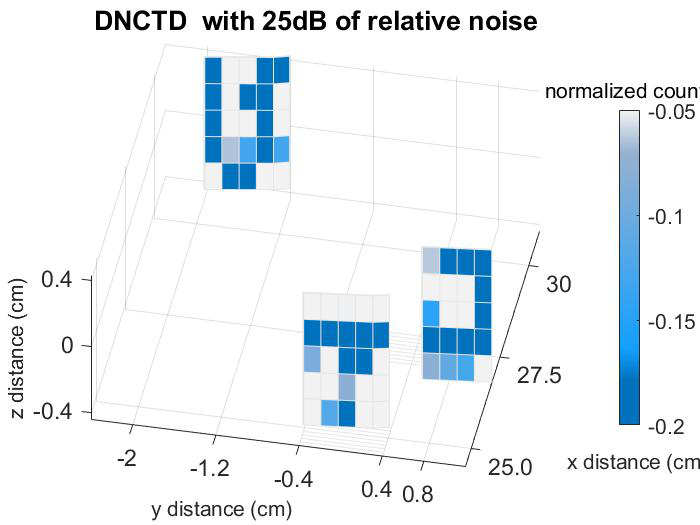}
  \caption{}
  \label{fig:NCTD Noise}
\end{subfigure}\newline
\begin{subfigure}{0.49\columnwidth}
  \centering
  \includegraphics[width=.9\columnwidth]{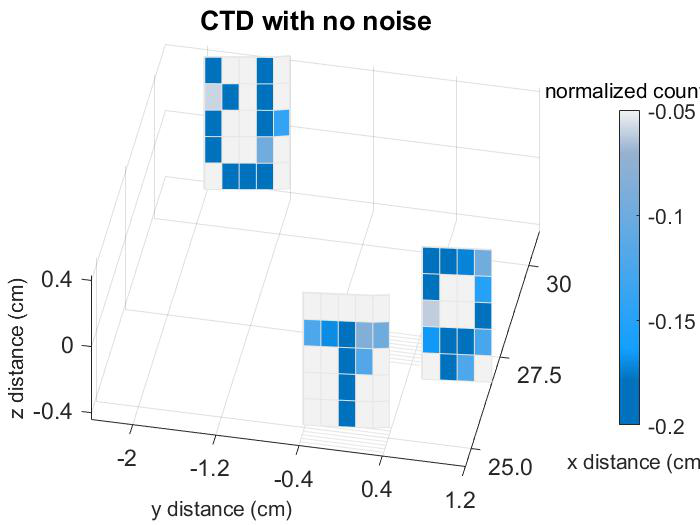}
  \caption{}
  \label{fig:NCTD Noise2}
\end{subfigure}
\begin{subfigure}{0.49\columnwidth}
  \centering
  \includegraphics[width=.9\columnwidth]{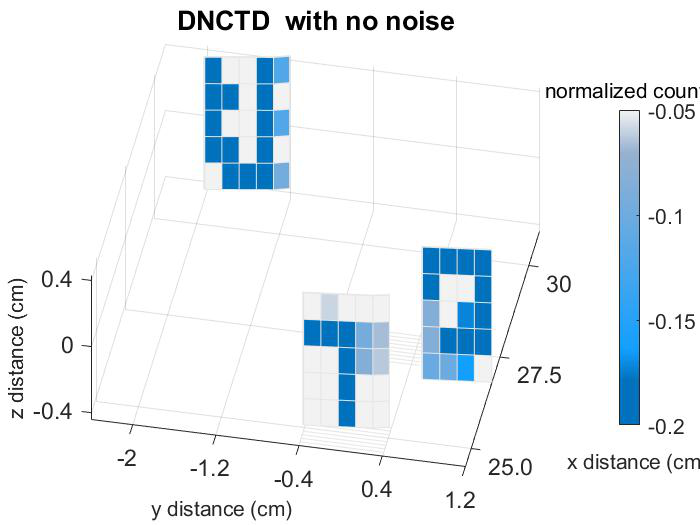}
  \caption{}
  \label{fig:NCTD Noise3}
\end{subfigure}\newline
\begin{subfigure}{0.49\columnwidth}
  \centering
  \includegraphics[width=0.9\columnwidth]{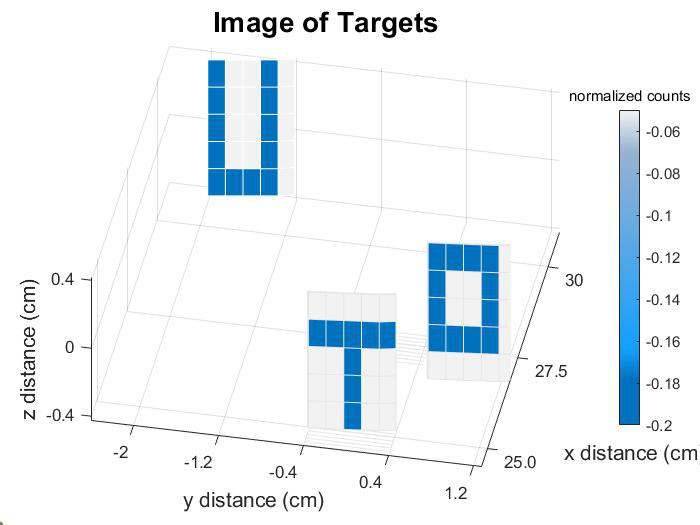}
  \label{fig:NCTD Noise4}
   \caption{}
\end{subfigure}
  \caption{ Scanning results for (a) CTD scheme with a 25dB noise background, (b) DNCTD scheme with 25dB of noise background, (c) CTD with 0dB noise background, (d) DNCTD scheme with 0dB noise background, (e) A numerical simulation of ideal imaging of all three letters. }
\end{figure}

Figures \ref{fig:CTD Noise},\ref{fig:NCTD Noise},\ref{fig:NCTD Noise2},\ref{fig:NCTD Noise3} depict the scanning results for normalized noise powers $0$db and $25$dB for both the CTD and DNCTD schemes. Three images ('U', 'O', and 'T') made of non-reflecting tape placed on a 1" mirror were scanned at different depths and positions. The intensity of each pixel was measured by using either the probe singles counts (CTD) or coincidence counts (NCTD and DNCTD). The targets were angled as to be not perpendicular to the incident beam to demonstrate the finest depth resolution achievable by our system. In the absence of environmental noise, all three letters are clearly visible in both schemes. However, when the noise is increased well beyond the signal power (25dB), all three letters become completely indistinguishable in the CTD scheme. In comparison, all three letters remain clearly readable in the DNCTD scheme. In all cases where the letters are visible, the ranging resolution was $\pm 0.09$cm allowing us to not only determine the range of each letter but the difference between the close and far side of each letter.

\section{Conclusion}
In conclusion, we were able to demonstrate a significant enhancement in target distinguishability from background noise of a phase-insensitive target-detection scheme. This was achieved by using the non-classical temporal correlations in SPDC photon pairs in conjunction with the relative reduction in noise power in the coincidence window resulting from the non-local dispersion cancellation. We quantified the performance by comparing the SNR of the CTD and DNCTD schemes and found a maximum enhancement of $43.1$dB. We incorporated this scheme with scanning and collection optics to image a target in a noisy environment far beyond what is classically possible at the same probe flux. 
Our scheme is still limited by a maximum noise flux of around $3.2\times 10^6$counts/second or $3.2$MHz above which the detector will saturate. However, due to the pulsed nature of the noise source, this is around 3 times greater than that previously recorded  in a temporal correlation based LiDAR setup in the CW regime \cite{liu2019enhancing}. Moreover, it is orders of magnitude larger than the $6.9$kHz noise rate recorded in \cite{tolt2018peak} and $0.4$kHz noise rate recorded in \cite{li2021single}. An interesting future study would be to investigate how other non-local effects could be used to further increase the noise-resilience of our target detection scheme. 

\bibliography{bibliography.bib}
\end{document}